# Nonlinear Non-Hermitian Skin Effect and Skin Solitons in Temporal Photonic Feedforward Lattices

Shulin Wang[1,2], Bing Wang[1,*], Chenyu Liu[1], Chengzhi Qin[1], Lange Zhao[1], Weiwei Liu[1], Stefano Longhi[3,4,†], and Peixiang Lu[1,5,§]

[1]*School of Physics and Wuhan National Laboratory for Optoelectronics, Huazhong University of Science and Technology, Wuhan, 430074, China.*

[2]*School of Physics, Southeast University, Nanjing 211189, China.*

[3]*Dipartimento di Fisica, Politecnico di Milano, Piazza Leonardo da Vinci 32, I-20133 Milano, Italy.*

[4]*IFISC (UIB-CSIC), Instituto de Fisica Interdisciplinar y Sistemas Complejos, E-07122 Palma de Mallorca, Spain.*

[5]*Hubei Key Laboratory of Optical Information and Pattern Recognition, Wuhan Institute of Technology, Wuhan 430205, China.*

Corresponding authors:

[*]wangbing@hust.edu.cn

[†]stefano.longhi@polimi.it

[§]lupeixiang@hust.edu.cn

**Abstract**

Here we report the experimental demonstration of the nonlinear non-Hermitian skin effect (NHSE) in an effective Kerr nonlinear temporal photonic lattice, where the high-power requirements and lack of tunability intrinsic to optical materials are overcome by an artificial nonlinearity arising from optoelectronic feedforward. Thanks to Kerr self-trapping, the nonlinear NHSE is demonstrated to possess much better localization strength and robustness at the preferred boundary compared to the linear case. Away from the preferred boundary, Kerr self-trapping can even inhibit NHSE-induced transport and form stable skin solitons. Harnessing the nonlinearity-controlled NHSE, we judiciously design an optical router with a flexibly tuned output port. Our findings promise great applications in robust signal transmission, routing, and processing.

Non-Hermitian (NH) physics has garnered significant attention over the past two decades due to its distinctive properties, which stand in stark contrast to those of traditional Hermitian systems [1-4]. Unlike Hermitian systems, which are characterized by real eigenvalue spectra and time-reversible dynamics, NH systems can exhibit complex eigenvalues, leading to phenomena such as nonreciprocal wave propagation, exceptional points, and the non-Hermitian skin effect (NHSE). These unique behaviors have opened up new avenues of research and potential applications, particularly in areas such as photonics, quantum mechanics, and condensed matter physics. The NHSE is perhaps one of the most intriguing phenomena in this exciting field [5-9]. The NHSE describes the localized behaviors of eigenstates near preferred edges in a lattice with asymmetric hoppings and open boundaries, which remarkably differs from the extended Bloch waves in the Hermitian case. Because of the NHSE, the traditional bulk-boundary correspondence fails to predict the topological transition and needs to be revised, which has recently evoked the rapid development of NH topological physics. Beyond the above physical innovations, the intrinsic localization of the NHSE and its sensitivity on the boundary condition host important applications in energy harvesting [10] and quantum sensing [11, 12]. An optical system is considered as a fantastic platform to observe the NHSE and implement its application [10, 13-15]. For example, by using the temporal quantum-walk lattice arising from two coupled fiber loops with gain and loss, the NHSE under the domain-wall boundary condition was experimentally realized and applied for topological light funneling [10].

Despite the above successes, nearly all previous studies of the NHSE were performed in the linear regime. The interplay of nonlinearity and skin localization provides a promising frontier of research, combining some key concepts of nonlinear physics such as solitons and NH topology of linear lattice systems, paving the way toward new ways for wave localization control and management. Until quite recently, only a few works shed light on the nonlinear NHSE [16-22], predicting the competition between nonlinear localization in the bulk and skin effect, and the existence and stability of nonlinear skin solitons both at system edges and bulk [22]. However, research on the nonlinear NHSE and its potential applications is still in its early stages, with experimental demonstrations of the nonlinear NHSE and skin solitons yet to be realized.

In this work we report on the first experimental demonstration of the Kerr nonlinear NHSE and skin solitons in a nonlinear optical system employing a synthetic Kerr nonlinear temporal lattice. To overcome high-power requirements and the lack of tunable nonlinearity in typical nonlinear optical setups [23-27], we introduce artificial tunable nonlinearity generated from opto-electronic feedforward. The skin solitons both at the edges and in the bulk at high nonlinearities are demonstrated, showing enhanced robustness under phase disorder by comparing with their linear counterpart. Inspired by the nonlinearity-controlled NHSE, we design a light router with its output port flexibly tuned by nonlinear strength. Our study offers a framework to explore and harness the interplay of NH topology and the nonlinear effect in unprecedented ways, with potential applications in robust light signal transmission, routing, and processing.

A synthetic temporal lattice can be established from two coupled fiber loops. As diagrammed in Fig. 1(a), the two loops have slightly different lengths and are bridged by a variable optical coupler (VOC). The average time delay and time delay difference of the two loops are $T$ and $2\Delta T$, respectively. When an optical pulse passes through the VOC, it splits into two parts and enters different loops. The parts circulating in the short and long loops relatively acquire a time advance $-\Delta T$ and a time delay $\Delta T$ compared to the averaging circulation time $T$, as shown in Fig. 1(b) where $m$ is the circulation number in a single loop. Along the real time axis $t$, by encoding the time slots of light pulses within an identical circulation into integers, we can obtain a series of discrete sites $n = (t-mT)/\Delta T$ and thus construct a lattice model in the time domain [10, 15, 25-33]. Moreover, the circulation number acts as the evolution dimension $m$ of the lattice. The pulse circulations in the short and long loops correspond to the leftward and rightward hoppings within the lattice. To realize the nonreciprocal hoppings and open boundary condition (OBC) essential for the NHSE, we introduce loss and gain modulation and dynamically-controlled coupling into the two loops. The gain factors in leftward and rightward hoppings are denoted by $e^{-\gamma}$ and $e^{\gamma}$, respectively. The sets of available sites under the OBC are chosen as $\{-N, -(N-2), ..., N-2, N\}$ and $\{-(N+1), -(N-1), ..., N-1, N+1\}$ in the even and odd steps, where $N$ is a positive even integer. The details of the experimental setup and OBC construction are shown in Supplemental Material Secs. I and II [34].

By incorporating optoelectronic feedforward circuits into the two loops, we can introduce effective Kerr nonlinearity into the temporal lattice. An optoelectronic feedforward circuit mainly consists of an optical coupler (OC), a single-mode fiber (SMF), a photodiode (PD), an amplifier (AMP), and a phase modulator (PM). The OC separates a minor part of the light pulse train out of the fiber loop. The PD and subsequent AMP transfer the optical pulse sequence into an electric signal with considerable amplitude. By applying the electric signal on the PM, the optical pulse train will acquire an additional phase modulation. Such a modulation is directly proportional to the intensity profile of the light signal, thus resembling the phase shift induced by the Kerr nonlinearity. Quantitatively, the effective Kerr nonlinear coefficient in our scheme can be calculated as $\kappa = \pi G_e R G_o I/V_\pi$, where $I$ is the light intensity before the OC, $G_o$ and $G_e$ are the optical power gain factor and voltage gain factor of the feedforward circuit, $R$ is the voltage responsivity of the PD, and $V_\pi$ is the half-wave voltage of the PM. Note that feedforward is more convenient for realizing synchronized modulation and emulating the Kerr effect, in comparison to feedback [35-38]. Most importantly, the effective Kerr coefficient can be readily modified by changing the optical or electric losses of the devices or gain factor of the AMP. In addition, the nonlinear loop circuit is operated at certain milliwatt light power and gets rid of the natural optical nonlinearity. These peculiar properties are quite desirable in nonlinear optics and facilitate the experimental exploration of nonlinear physics.

The pulse dynamics in the nonlinear NH temporal lattice realizes a one-dimensional nonlinear quantum walk, which reads [10, 15, 25-33]

$$u_n^{m+1} = e^{-\gamma} U_{n+1}^m e^{i\chi|e^{-\gamma}U_{n+1}^m|^2}, \quad v_n^{m+1} = e^{\gamma} V_{n-1}^m e^{i\chi|e^{\gamma}V_{n-1}^m|^2}, \tag{1}$$

with $U_{n+1}^m = \cos(\beta)u_{n+1}^m + i\sin(\beta)v_{n+1}^m$ and $V_{n-1}^m = \cos(\beta)v_{n-1}^m + i\sin(\beta)u_{n-1}^m$. Here, $u_n^m$ and $v_n^m$ represent the pulse amplitudes in the short and long loops normalized by the square root of the total intensity of incident pulses $I_0$, and $\beta$ determines the coupling ratio of the VOC in the form of $\sin^2(\beta)/\cos^2(\beta)$. The variable $\chi$ denotes the Kerr coefficient defined for the normalized pulse amplitudes, which reads $\chi = \pi G_e R G_o I_0/V_\pi$. We focus our study mostly on a lattice with the OBC. In this case, at the left and right edges, the pulse evolution (1) from odd to even steps needs to be remodified by setting $\beta = \pi/2$ (Supplemental Material Sec. II [34]), whereas from even to odd steps the pulse dynamics is identical to the bulk one. As shown in Supplemental Material Sec. III [34], the nonlinear quantum walk described by Eq. (1) realizes the nonlinear Hatano-Nelson model [16, 20, 22] when the coupling angle $\beta$ is close to $\pi/2$ and the nonlinearity is weak. Hence, we expect our model sustain nonlinear skin solitons and to show competition between skin localization and nonlinear localization in the bulk [20, 22].

To calculate the nonlinear OBC eigenvalues and eigenmodes, we derive the two-step propagator $\hat{U}$ from Eq. (1), which is denoted by the Hadamard product of a linear part $\hat{U}_L$ and a nonlinear part $\hat{U}_{NL}$, i.e., $\hat{U} = \hat{U}_L \circ \hat{U}_{NL}$ (see Supplemental Material Sec. IV [34]). With $|\psi(m)\rangle = (v_{-N}^m, u_{-N}^m, v_{-(N-2)}^m, u_{-(N-2)}^m, ...,$ $v_{N-2}^m$, $u_{N-2}^m$, $v_N^m$, $u_N^m)^T$ being the vector of wave amplitudes, the two-step evolution reads $|\psi(m+2)\rangle = \hat{U}|\psi(m)\rangle$. We can introduce the nonlinear eigenmodes $|\Psi\rangle$ and corresponding nonlinear eigenvalues, i.e., the propagation constant $\theta$, via the nonlinear eigen equation

$$\hat{U}|\Psi\rangle = e^{i\theta}|\Psi\rangle, \qquad (2)$$

where $|\Psi\rangle = (v_{-N}, u_{-N}, v_{-(N-2)}, u_{-(N-2)}, ..., v_{N-2}, u_{N-2}, v_N, u_N)^T$. Because the intensity profile of the eigenstate can inversely determine the phases in $\hat{U}_{NL}$, the eigenmode is a self-consistent solution of a nonlinear eigen equation. Here we utilize the self-consistent iteration algorithm to numerically calculate the eigenmode [39] (Supplemental Material Sec. V [34]). We mention that, via the non-unitary gauge transformation $u_n^m \to u_n^m \exp(\gamma n)$ and $v_n^m \to v_n^m \exp(\gamma n)$, the NH terms can be removed from Eq. (1), and the dynamics becomes conservative which implies that the nonlinear eigenvalue $\theta$ can be real under the OBC. To reveal how the Kerr nonlinearity influences the localization strength of the eigenmode, we calculate the inverse participation ratio (IPR) of the mode profile IPR = $\Sigma_n(|u_n|^2+|v_n|^2)^2/[\Sigma_n(|u_n|^2+|v_n|^2)]^2$ [29]. The IPR varies within (0, 1], and a larger IPR reflects a higher degree of localization of the eigenmode. We also mention that, in the linear or weakly nonlinear regime, using the bulk band structure can efficiently describe the NHSE dynamics within the bulk, which manifests as a unidirectional amplifying transport (Supplemental Material Sec. VI [34]).

To elucidate the influence of Kerr nonlinearity on the NHSE, we comprehensively compare the mode profiles, propagation constants, and IPR of linear and nonlinear skin modes in Fig. 2. The lattice sites range from $n = -10$ to 10, and the gain and loss parameter and coupling angle are set as $\gamma = 0.2$ and $\beta = \pi/3$. In the linear case [Fig. 2(a)], all skin modes localize near the right edge $n = 10$, and their propagation constants are within the real part of bulk band structure. The above propagation constants are entirely real, which is starkly in contrast to the complex bulk band structure (Supplemental Material Sec. VII [34]). Such localized behaviors of the eigenmodes and deviation of the spectrum evidently demonstrate the existence of the NHSE. However, these skin modes widely distribute from $n = 0$ to 10, and the average IPR is calculated as 0.35 based on the averaged intensity profile of all skin modes, which indicates a relatively weak localization. To confirm the above analysis, we simulate and measure the propagation of the most localized skin mode, i.e., the sixth mode labeled in blue in Fig. 2(a). In experiment, the skin mode is prepared by applying intensity, phase, and coupling modulations during single pulse diffraction. As shown in Figs. 2(b) and 2(c), the experimental result coincides quite well with the simulated one. The skin mode stably evolves with an unchanged profile and localizes near the right boundary.

The nonlinear skin modes can exhibit different properties compared with the linear ones [16, 20, 22]. Using the self-consistent iteration algorithm, we can obtain two stable nonlinear skin modes at the right edge, i.e., the modes 1 and 2, in which the optical energy is mainly concentrated in the long and short loops, respectively. Such modes can be regarded as edge skin solitons [22], where the skin effect and self-trapping cooperatively offer strong localization at the right boundary. As depicted in Fig. 2(d), the mode profile is remarkably compressed compared to the linear skin mode. This is because the Kerr nonlinear phase can form a nonlinear waveguide [40, 41], thereby further confining the localized linear skin mode. More interestingly, the nonlinear waveguide can lift the propagation constant into the gap of the bulk band structure. Akin to the topological edge state, the nonlinear skin mode is hard to couple with bulk modes and hence shows highly localized behaviors [26, 42, 43]. The IPR enhancement also supports the above analysis. However, at the edges of the band gap (corresponding to very weak or strong nonlinearity), we fail to search for a stable skin mode with high localization considering the closeness to the bulk bands [42, 43]. Figures 2(e) and 2(f) show the simulated and measured eigenmode evolutions with $\chi = 0.4\pi$, where the pulses are well confined within the two rightmost sites and periodically revive after each of two steps. For the nonlinear skin mode 2, the mode profile becomes extremely localized and even distributes within nearly a single site, as illustrated in Figs. 2(g)-2(i). This is because at $n = 10$ the sublattice corresponding to the short loop is exactly the right boundary of the lattice, and the reflection at boundary can further confine the mode profile.

To examine the concerned robustness of the NHSE [10, 44], we introduce transverse random phase noise $\alpha(n)$ onto the skin mode evolutions shown in Fig. 2. The phase noise ranges from $-\Delta\alpha/2$ to $\Delta\alpha/2$ with $\Delta\alpha$ being the noise strength. Figures 3(a)-3(c) illustrate the experimental evolutions of the sixth linear skin mode and the nonlinear skin modes 1 and 2, in which the noise strength is chosen as $\Delta\alpha = 0.25\pi$. In the linear case, the skin mode still localizes around the right edge. Nevertheless, its intensity pattern is severely distorted by the phase disorder, accompanied with irregular energy amplification and attenuation during evolution. For the nonlinear skin mode 1, only a small amount of distortion is observed. Remarkably, for the nonlinear skin mode 2, its evolution is even immune to the disorder. This is because the induced nonlinear waveguide shifts the propagation constant of the nonlinear skin mode 2 to the center of the bulk band gap, thereby giving rise to very strong robustness.

The suppressed intensity distortion at the right edge $n = 10$ [Figs. 3(d)-3(f)] also confirms the enhanced robustness of the nonlinear NHSE.

In the bulk or at the left edge, the NHSE manifests as a rightward transport instead of localization, and competition with the Kerr self-trapping arises. Such a competition can lead to a total suppression of the NHSE and thus yields bulk skin solitons away from the right edge [22]. As shown in Fig. 4(a), the bulk mode also has a highly confined albeit asymmetric profile. As an illustration with $\chi = 0.5\pi$, a stable mode propagation can be observed in the bulk [Fig. 4(b)], evidently demonstrating the nonlinear suppression of the NHSE. Because of the translation invariance of the lattice, the bulk modes with the same characteristics can also be formed around other sites in the bulk. More interestingly, we can even obtain a localized eigenmode at the left edge [Fig. 4(c)]. The experiment [Fig. 4(d)] clearly supports the validity of the nonlinear left-edge mode. For comparison, we plot in Figs. 4(e) and 4(f) the eigenvalue spectra and IPR varying with $\chi$ for the nonlinear bulk and edge modes. The threshold of nonlinearity for the formation of the bulk skin soliton is clearly raised compared to the nonlinear skin mode at the right edge. Owing to the edge reflection, the threshold for the existence of the left-edge skin soliton is further increased, with the mode profile slightly compressed.

Using the mode formation around different sites, here we ingeniously design a nonlinear optical router with its output position flexibly tuned by the nonlinear strength. Specifically, a single weak pulse is adopted to excite a site away from the right edge. For a small nonlinear coefficient $\chi$, the components possessing maximum group velocity first dominate the pulse dynamics (see Supplemental Material Sec. VI). The pulses will move toward the right boundary accompanied with exponentially increasing intensities. When approaching the boundary, the amplified pulses yield strong enough nonlinearity and form nonlinear skin modes. A robust routing is hence generated, as presented in Figs. 5(a) and 5(c). For a large enough $\chi$, the nonlinear eigenmode can be formed as soon as the pulse is injected, leading to a routing around the initial position [Figs. 5(b) and 5(d)]. For a moderate $\chi$, a routing between the incident position and right boundary can be achieved with its output port tuned by $\chi$, as indicated in the box area of Fig. 5(e). The output pulse train moves leftward to the initial position for increasing $\chi$. The intensity profiles manifest moderately strong localization due to the considerable IPR. More details are presented in Supplemental Material Sec. VIII [34]. It should be mentioned that the left-edge skin soliton cannot be directly excited by the single-site bulk excitation since the hopping toward the left is always attenuated in the non-Hermitian system. Apart from the effective Kerr coefficient, the routing effect can be equivalently realized by varying the intensity of the incident pulses because the nonlinear phase is determined by the product of the Kerr coefficient and intensity, as implied in Eq. (1).

In conclusion, we reported on the first experimental demonstration of the nonlinear NHSE and skin solitons, unveiling the intriguing competition or cooperation of the NHSE and nonlinear self-trapping. Our experiment utilizes temporal quantum-walk optical lattices incorporated with optoelectronic feedforward, providing a highly alternative platform to realize the Kerr nonlinearity with a high degree of tunability. At the preferred boundary, i.e., the right boundary here, the linear NHSE and nonlinear self-trapping cooperate, owing to the formation of a nonlinear waveguide, yielding extremely localized nonlinear skin modes (edge skin solitons). Furthermore, under phase disorder, edge skin solitons are demonstrated to have highly enhanced robustness compared to the linear case. More surprisingly, in the bulk or at the left edge, the NHSE can be entirely suppressed as the Kerr self-trapping is strong enough to overcome the NHSE-induced transport, giving rise to the formation of bulk or left-edge skin solitons. Our work pushes the celebrated NHSE into the nonlinear regime, providing a bridge between the NH topology and the nonlinear physics. Our results could significantly influence practical applications, creating a unique pathway for applying nonlinear NHSE physics to robust light signal transmission, routing, and processing. In addition, the low optical power requirement and high tunability of our mimicking Kerr nonlinearity may profoundly boost the development of basic nonlinear devices for optical computation, including couplers [45, 46], switches [47, 48], and interferometers [49, 50].

This work is supported by the National Natural Science Foundation of China (No. 12374305, No. 62305122, No. 12474381 and No. 12021004). S.L. acknowledges the Spanish Agencia Estatal de Investigacion (MDM-2017-0711).

FIG. 1. (a) Schematic of two coupled fiber loops incorporated with optoelectronic feedforward. (b) Construction principle of temporal quantum-walk lattice based on time multiplexing of a light pulse.

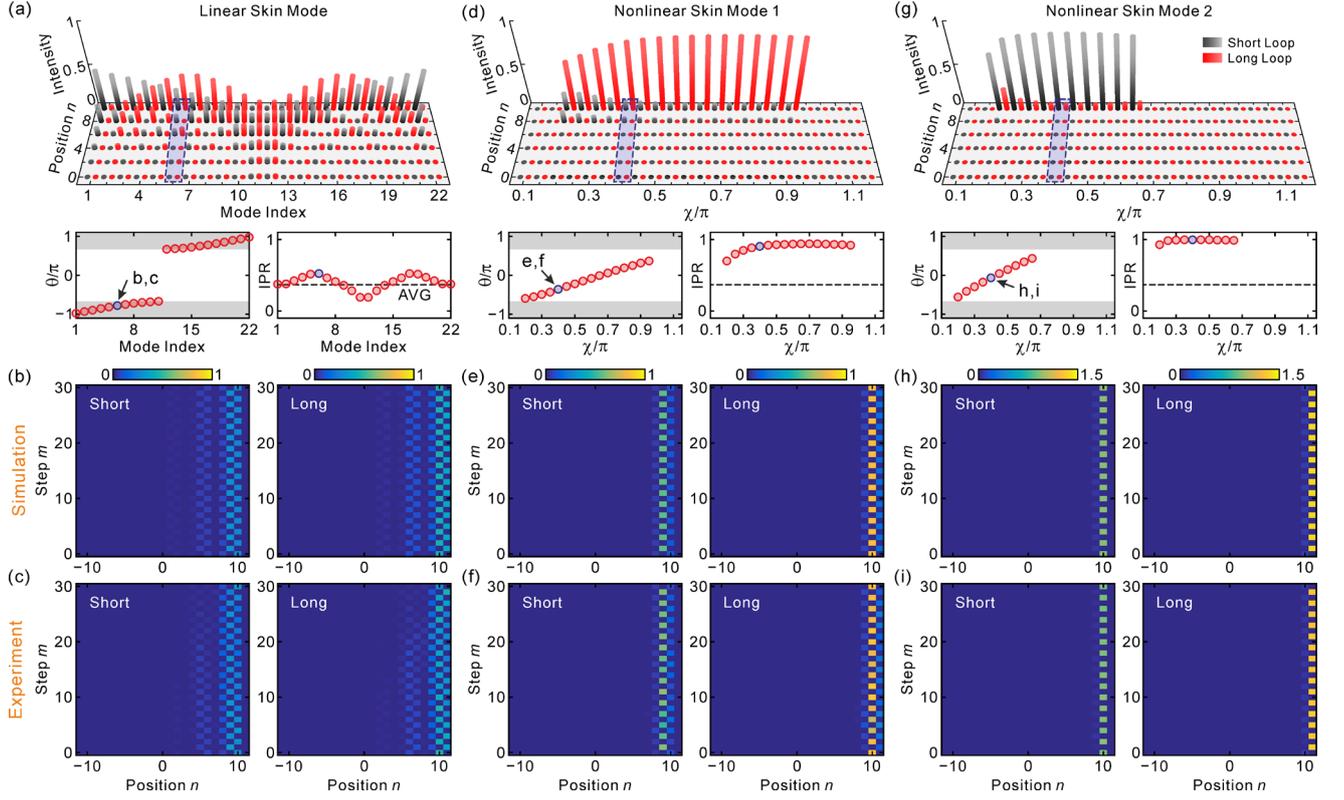

FIG. 2. (a) Mode profiles, eigenvalue spectrum, and IPR of linear skin modes. The black and red bars in the mode profile subplot correspond to short and long loops, respectively. The gray zones in the spectrum indicate the real parts of the bulk band structure, and the circles correspond to the skin modes. The dashed line in the IPR subplot represents the IPR of the average intensity profile of all skin modes. (b),(c) Simulated and measured intensity evolutions of the sixth linear skin mode. (d) Mode profiles, eigenvalue spectrum, and IPR of nonlinear skin mode 1 versus Kerr coefficient $\chi$. (e),(f) Simulated and measured intensity evolutions of nonlinear skin mode 1 for $\chi = 0.4\pi$. (g)-(i) correspond to the nonlinear skin mode 2.

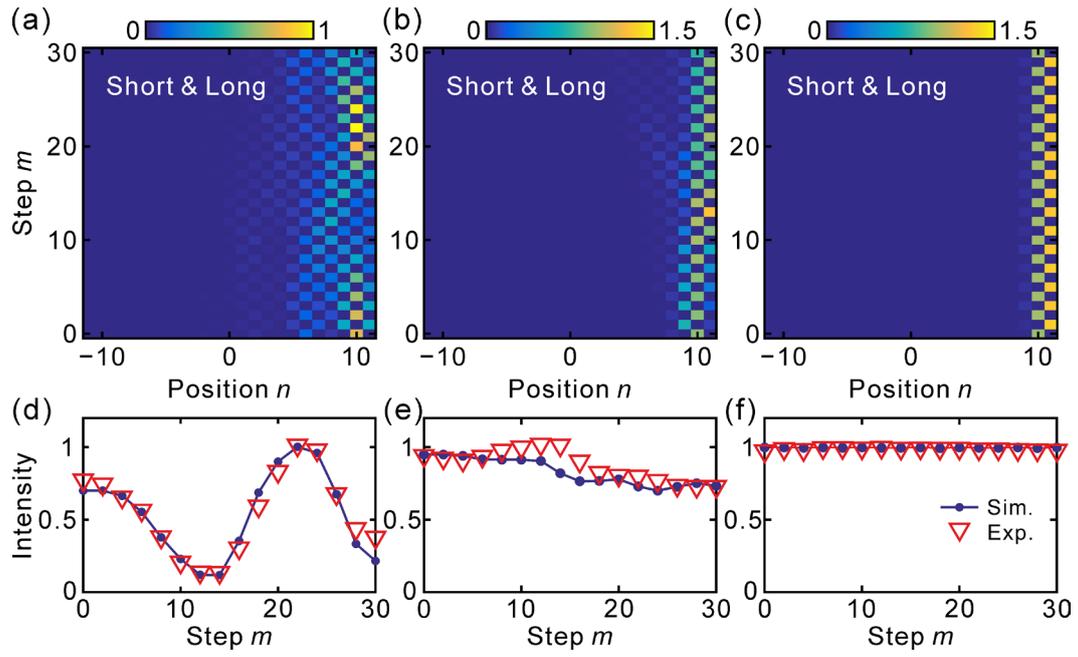

FIG. 3. (a)-(c) Measured pulse intensity evolutions of the sixth linear skin mode and nonlinear skin modes 1 and 2 under phase disorder ($\Delta\alpha = 0.25\pi$). (d)-(f) Corresponding pulse intensity evolutions at $n = 10$. The blue dots and red triangles denote the simulated and measured results, respectively.

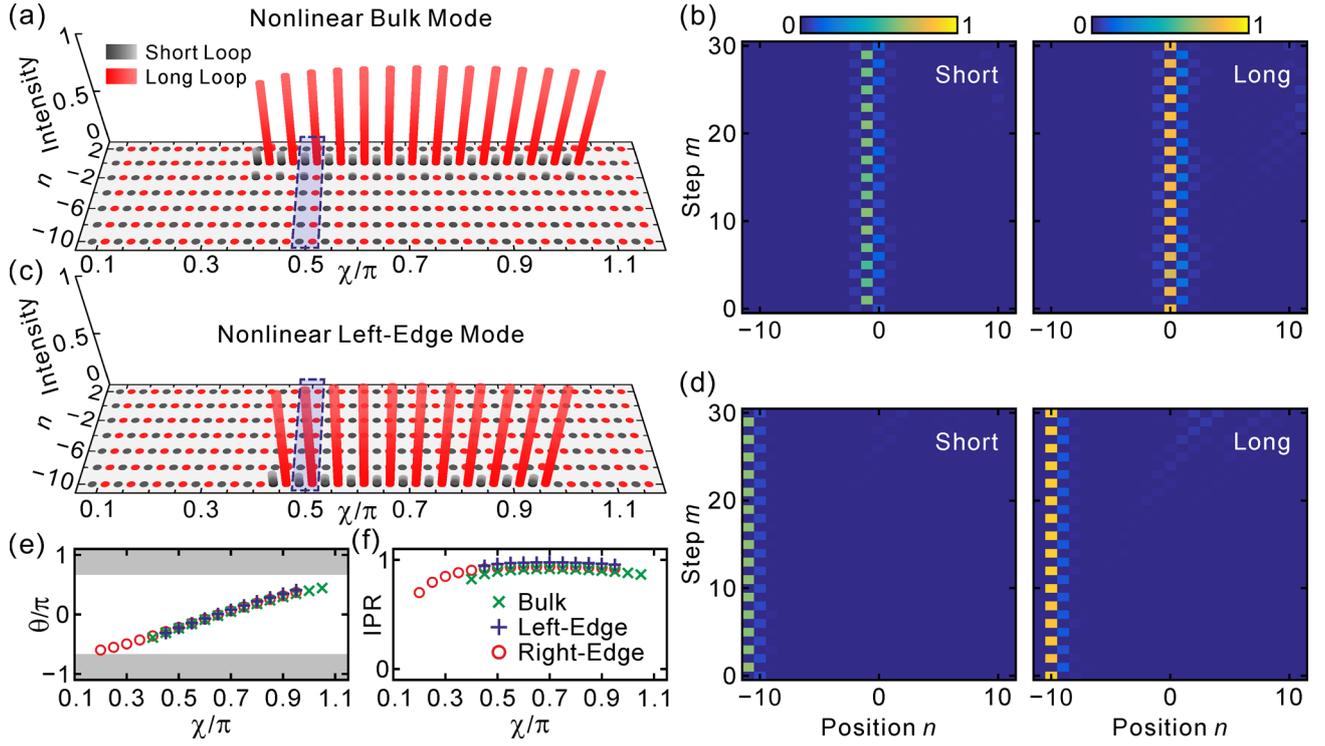

FIG. 4. (a) Mode profile varying with $\chi$ for the nonlinear bulk mode. (b) Measured intensity evolution of the nonlinear bulk mode at $\chi = 0.5\pi$. (c),(d) correspond to the nonlinear left-edge mode. (e), (f) Eigenvalue spectra and IPR varying with $\chi$ for the three nonlinear modes concentrated in the long loop.

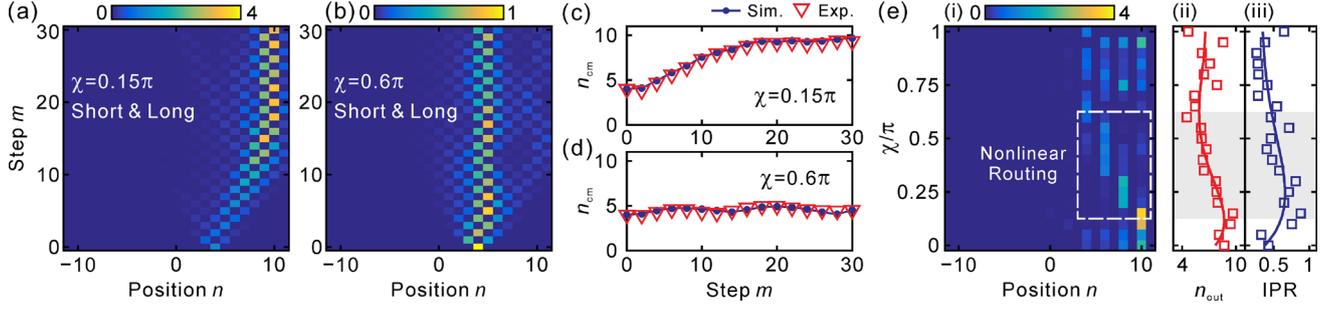

FIG. 5. (a),(b) Measured pulse intensity evolutions for single-site excitation at $\chi = 0.15\pi$ and $0.6\pi$. The pulse is injected into the long loop at $n = 4$. (c),(d) Corresponding center of mass $n_{cm}$. (e) Simulated intensity distribution (i), center of mass $n_{out}$ (ii), and IPR (iii) of the pulse sequence at $m = 30$ versus Kerr coefficient $\chi$. In (ii) and (iii), the squares and solid curves denote the simulated and fitting results.